\documentclass[%
aip,
pof,
preprint,
]{revtex4-1}

\usepackage{graphicx}%
\usepackage{amssymb}%
\usepackage{xspace}%

\newcommand{\micron}{$\mu$m\xspace}     
\newcommand{\us}{$\mu$s\xspace}         
\newcommand{\Celsius}{$^\circ$C\xspace} 

\begin{document}

\title{Microbubble formation and pinch-off scaling exponent in flow-focusing devices}

\author{Wim van Hoeve} \affiliation{Physics of Fluids, Faculty of Science and Technology, and MESA$^+$ Institute for Nanotechnology, University of Twente, P.O.~Box 217, 7500 AE Enschede, The Netherlands}
\author{Benjamin Dollet} \affiliation{Institut de Physique de Rennes, UMR UR1-CNRS 6251, Universit\'e de Rennes 1, Campus de Beaulieu, B\^atiment 11A, F-35042 Rennes Cedex, France}
\author{Michel Versluis}
\author{Detlef Lohse} \affiliation{Physics of Fluids, Faculty of Science and Technology, and MESA$^+$ Institute for Nanotechnology, University of Twente, P.O.~Box 217, 7500 AE Enschede, The Netherlands}



\date{25 February 2011}

\begin{abstract}
We investigate the gas jet breakup and the resulting microbubble formation in a microfluidic flow-focusing device using ultra high-speed imaging at 1 million frames/s. In recent experiments [Dollet \emph{et al.}, \href{http://dx.doi.org/10.1103/PhysRevLett.100.034504}{Phys.\ Rev.\ Lett.}\ \textbf{100}, 034504 (2008)] it was found that in the final stage of the collapse the radius of the neck scales with time with a $1/3$ power-law exponent, which suggested that gas inertia and the Bernoulli suction effect become important. Here, ultra high-speed imaging was used to capture the complete bubble contour and quantify the gas flow through the neck. It revealed that the resulting decrease in pressure, due to Bernoulli suction, is too low to account for an accelerated pinch-off. The high temporal resolution images enable us to approach the final moment of pinch-off to within 1\,\us.
We observe that the final moment of bubble pinch-off is characterized by a scaling exponent of $0.41\,\pm\,0.01$. This exponent is approximately $2/5$, which can be derived, based on the observation that during the collapse the neck becomes \emph{less slender}, due to the exclusive driving through liquid inertia.
\end{abstract}

\maketitle

\section{Introduction}
Liquid droplet pinch-off in ambient air or gas bubble pinch-off in ambient liquid can mathematically be seen as a singularity, both in space and time.\cite{Eggers1997,Eggers2008}
The process that leads to such a singularity has been widely studied in recent years\cite{ChenSteen1997,Day1998,Ganan-Calvo2001,Leppinen2003,Burton2005,Gordillo2005,Ganan-Calvo2006,Gordillo2006,Bergmann2006,Keim2006,Thoroddsen2007,Gordillo2008,Gekle2009PRE} and is of major importance in an increasing number of medical and industrial applications.
Examples of this are the precise formation and deposition of droplets on a substrate using inkjet technology,\cite{Wijshoff2010} or for the production of medical microbubbles used in targeted drug delivery.\cite{Ferrara2007,Ferrara2009}

For the pinch-off of liquid in gas, the dynamics close to pinch-off exhibit self-similar behavior, which implies that the local shape of the neck is not influenced by its initial conditions.
The radius of the neck goes to zero following a universal scaling behavior with $r_0 \propto \tau^\alpha$, where $\tau$ represents the time remaining until pinch-off and $\alpha$ the power law scaling exponent.\cite{Eggers1997}
The scaling exponent $\alpha$ is a signature of the physical mechanisms that drive the pinch-off.
The formation and pinch-off of a low-viscosity liquid droplet in air is described by a balance between surface tension and inertia, resulting in a $2/3$ scaling exponent.\cite{ChenSteen1997,Day1998,Leppinen2003,Burton2004,Eggers2008,Gonzalez2009}

The inverted problem of the collapse of a gaseous thread in a liquid is, however, completely different. Initially, a simple power law was predicted based on a purely liquid inertia driven collapse giving rise to a $1/2$ scaling exponent.\cite{Longuet-Higgins1991,Oguz1993,Burton2005} However, many groups report power law scaling exponents that are slightly larger than $1/2$.\cite{Gordillo2005,Bergmann2006,Keim2006,Thoroddsen2007,Gordillo2008,Bolanos-Jimenez2008,Bolanos-Jimenez2009} In recent work of Eggers \emph{et al.}\cite{Eggers2007} and Gekle \emph{et al.}\cite{Gekle2009PRE} it was demonstrated that a coupling between the radial and axial length scale of the neck\cite{Gordillo2006} can explain these small variations in the scaling exponent. Based on a slender-body calculation it is found that $\alpha(\tau) = 1/2 + (-16 \ln{\tau})^{-1/2}$, where $\alpha$ slowly asymptotes to $1/2$ when approaching pinch-off.

In the work of Gordillo \emph{et al.}\cite{Gordillo2005,Gordillo2008} it has been shown that gas inertia, \emph{i.e.} Bernoulli suction, plays an important role in the bubble pinch-off. The increasing gas flow through the neck results in an accelerated collapse with $\alpha = 1/3$.\cite{Gordillo2008,Gekle2010}
It should be noted that the smaller the scaling exponent $\alpha$ the more rapidly the radius of the neck diminishes at the instant of pinch-off, since the speed of collapse $\dot{r}_0 \propto \alpha \tau^{\alpha - 1}$, where the overdot denotes the time derivative.

In the work of Dollet \emph{et al.}\cite{Dollet2008} microbubble formation in a microfluidic flow-focusing device was investigated.
A flow-focusing device comprises two co-flowing fluids, an inner gas and an outer liquid phase, that are focused into a narrow channel where bubble pinch-off occurs.
It was found that bubble formation in a square cross-sectional channel ($W\,\times\,H = 20\,$\micron$\,\times\,20\,$\micron) showed a similar collapse behavior giving a $1/3$ scaling exponent. In that paper it was suggested that this exponent reflects the influence of gas inertia.
However, this scaling exponent could not be conclusively ascribed to Bernoulli suction, due to a lack of spatial and temporal resolution at the neck in the final stages of pinch-off.

In this work we study the bubble formation for extremely fast bubble pinch-off in a microfluidic flow-focusing channel of square cross-section, using ultra high-speed imaging at 1\,Mfps.
The complete spatial structure of the bubble, including its neck, was captured.
This allowed us to not only investigate the effect of Bernoulli suction, but also the influence of the constituent radial and axial length scale length scales of the neck.

Here we find that the ultimate stage of microbubble pinch-off is purely \emph{liquid} inertia driven. In our system, the neck becomes \emph{less slender} when approaching the pinch-off, giving rise to an exponent $\alpha = 2/5$ over almost 2 decades, which is different as compared to the case of bubble pinch-off in the bulk as reported by Bergmann \emph{et al.},\cite{Bergmann2006} Thoroddsen \emph{et al.},\cite{Thoroddsen2007} and Gekle \emph{et al.},\cite{Gekle2010} among others.

\section{Experimental setup}
\begin{figure}[h!]
    \includegraphics[]{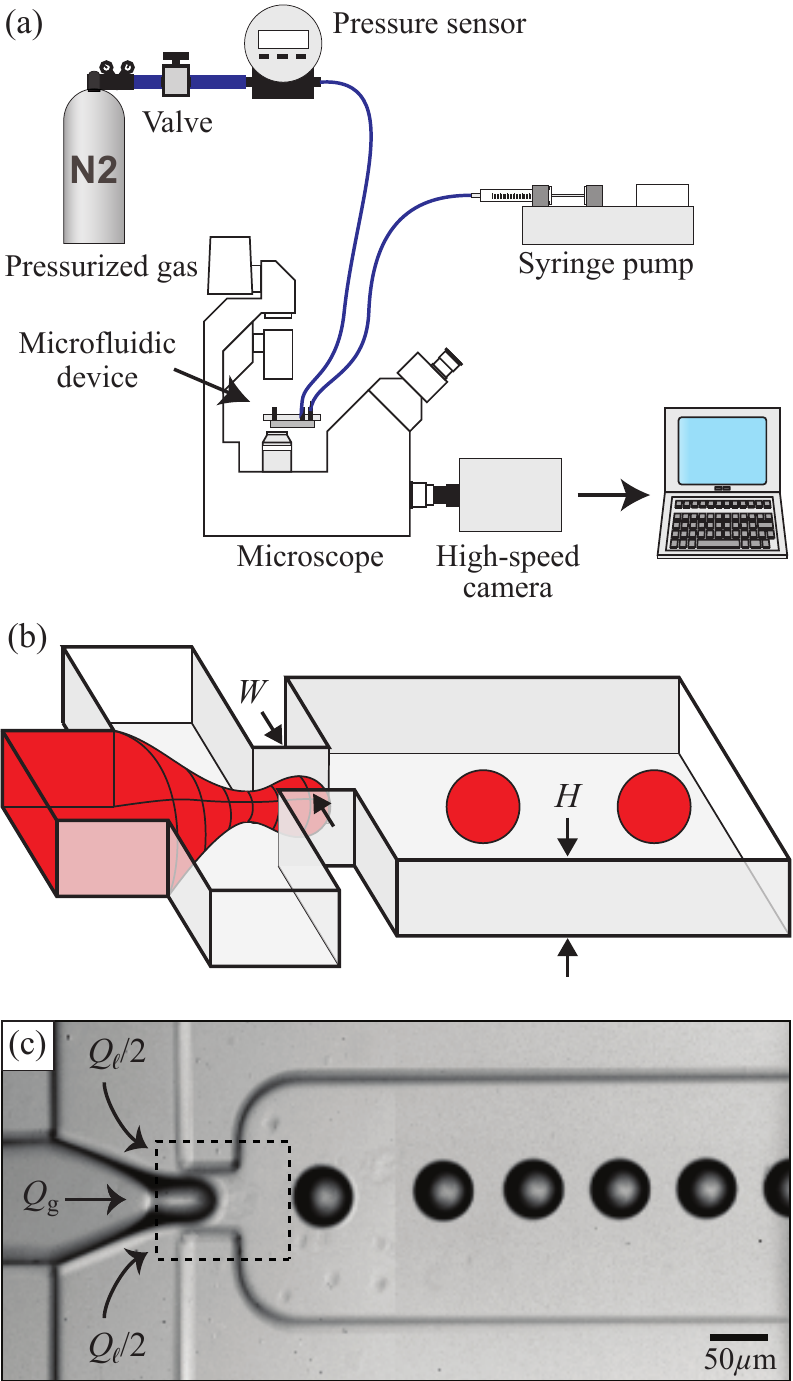}
    \caption{\label{Fig:ICTAM:flow_focusing_setup}(a) Schematic overview of the setup for the study of microbubble formation in microfluidic flow-focusing devices. A high-speed camera mounted to an inverted microscope is used to capture the final moment of microbubble pinch-off. Gas pressure was controlled by a pressure regulator connected to a sensor. The liquid flow rate was controlled by a high-precision syringe pump. (b) Schematic representation of a planar flow-focusing device with uniform channel height $H = 59$\,\micron and channel width $W = 60$\,\micron. (c) Snapshot of a high-speed recording. The outer liquid flow $Q_\ell$ forces the inner gas flow $Q_\mathrm{g}$ to enter a narrow channel (encircled by the dashed line) in which a microbubble is formed.}
\end{figure}
The experimental setup is shown in Fig.~\ref{Fig:ICTAM:flow_focusing_setup}a. The flow-focusing device is fabricated with a square cross-section channel geometry, with channel width $W = 60$\,\micron and height $H = 59$\,\micron{}, as depicted schematically in Fig.~\ref{Fig:ICTAM:flow_focusing_setup}, to ensure that the collapse occurs in the radial 3D collapse regime only.\cite{Dollet2008}
The device was produced using rapid prototyping techniques.\cite{Duffy1998} A homogeneous layer of negative photoresist (SU-8) is spin-coated on a silicon wafer. The thickness of the layer defines the channel height. A chrome mask (MESA$^+$ Institute for Nanotechnology, University of Twente, The Netherlands) is used in contact photolithography to imprint features with sizes down to 2\,\micron. After ultraviolet exposure a cross-linking reaction starts which rigidifies the photoresist that is exposed to the light. The photoresist that is not exposed is removed during development with isopropanol. What is left is a positive relief structure which can be used as a mold to imprint micron-sized channels in polydimethylsiloxane (PDMS) (Sylgard 184, Dow Corning). PDMS is a transparent polymer which is obtained by mixing two components, base and curing agent, in a 10:1 ratio in weight. The mixture is poured on the mold and cured in a 65\Celsius oven for 1 hour. The PDMS slab with imprinted microchannels is removed from the mold and then holes are punched in the PDMS. The PDMS slab is oxygen plasma-bonded (Harrick Plasma, Model PDC-002, Ithaca, NY, USA) to a glass cover plate of 1\,mm thickness to close the channels. Plasma bonding creates a non-reversible bond which can withstand pressures up to a few bars.\cite{Bhattacharya2005, Eddings2008} The oxygen plasma turns the PDMS channel walls temporarily hydrophilic which enhances fluid flow and  wetting of the channel walls. After closing the device, $1/16$ inch outer diameter Telfon tubing is connected to the inlet channels, through which gas and liquid is supplied.

Nitrogen gas is controlled by a regulator (Omega, PRG101-25) connected to a pressure sensor (Omega, DPG1000B-30G). The gas supply pressure was 12\,kPa. A 10\% (w/w) solution of dishwashing liquid (Dreft, Procter \& Gamble) in deionized water is flow-rate-controlled using a high precision syringe pump (Harvard Apparatus, PHD 2000, Holliston, MA, USA). The liquid, with density $\rho = 1000$\,kg/m$^3$, surface tension $\gamma = 35$\,mN/m, and viscosity $\eta = 1\,\mbox{mPa}\cdot$s, wets the channel walls. The liquid surfactant solution was supplied at a flow rate $Q_\ell = 185\,\mu{}$l/min. The Reynolds number $\textrm{Re} = \rho Q_\ell R / \eta W H \approx 26$, with nozzle radius $R = W H / \left( W + H \right) \approx 30$\,\micron, is low enough to guarantee that the flow is laminar.

The bubble formation process is imaged using an inverted microscope (Nikon Instruments, Eclipse TE2000-U, Melville, NY, USA) equipped with an extra long working distance objective with a cover glass correction collar (Nikon Instruments, ${60\times}$ Plan Fluor ELWD N.A.~0.70 W.D.~2.1--1.5\,mm, Melville, NY, USA) and an additional ${1.5\times}$ magnification lens. The system is operated in bright-field mode using high-intensity fiber illumination (Olympus, ILP-1, Zoeterwoude, The Netherlands).

To resolve the growth of the bubble and the extremely fast bubble pinch-off at the same time requires a high-speed camera that is capable of recording images at a high frame rate and at full resolution so that the field of view is sufficient to capture the entire bubble profile at sufficiently high spatial resolution.
These two criteria, \emph{i.e.}~a short interframe time (of the order of 1\,\us) and a sufficiently large field of view means a specialized ultra high-speed camera is required for this task.
Hence, we use the Shimadzu ultra high-speed camera (Shimadzu Corp., Hypervision HPV-1, Kyoto, Japan) to capture 100 consecutive images at a high temporal resolution of 1\,Mfps (equivalent to an interframe time of 1\,\us), exposure time of 0.5\,\us, field of view of 200\,\micron{}$\,{\times}\,$175\,\micron{}, and with a spatial resolution of 0.68\,\micron/pixel.

\begin{figure*}
    \includegraphics[width=14cm]{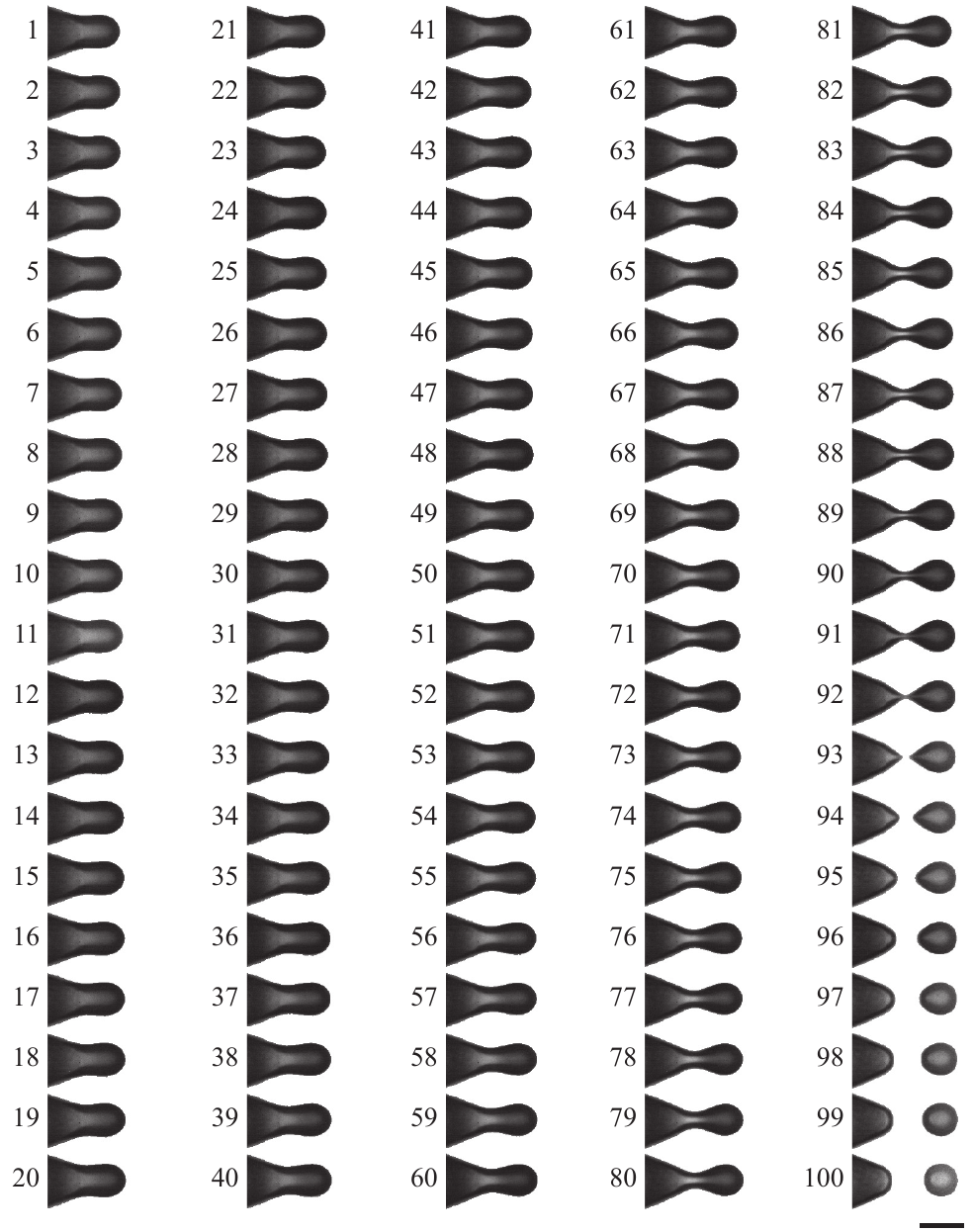}
    \caption{\label{Fig:ICTAM:timeseries}Time series showing the formation of a microbubble in a microfluidic flow-focusing device recorded at 1\,Mfps. The frame number is indicated at the left of each frame. For reasons of clarity, the background, including the channel structure, is subtracted. A detailed image that corresponds to frame 86 is represented in Fig.~\ref{Fig:ICTAM:coordinates}a. The camera's field of view is indicated by the dashed line in Fig.~\ref{Fig:ICTAM:flow_focusing_setup}c. The exposure time is 0.5\,\us{}. The scale bar in the lower right corner denotes 50\,\micron{}.}
\end{figure*}
\section{Results}
\subsection{Extracting the collapse curves}
In Fig.~\ref{Fig:ICTAM:timeseries} a time series of the formation of a microbubble is shown, where all images are background subtracted to improve the contrast. The first image (frame 1) shows the bubble almost completely blocking the narrow channel (\emph{cf.}~Fig.~\ref{Fig:ICTAM:flow_focusing_setup}c). This restricts the outer liquid flow and the liquid starts to squeeze the gas in the radial direction forming a neck.
The neck becomes smaller and smaller until final pinch-off, resulting in bubble detachment (frame 93).
The complete contour of the bubble is extracted from the recordings using image analysis algorithms in MATLAB (Mathworks Inc., Natick, MA, USA). In order for precise detection of the contour the images were resampled and bandpass filtered in the Fourier domain to achieve sub-pixel accuracy.
\begin{figure}
    \includegraphics[]{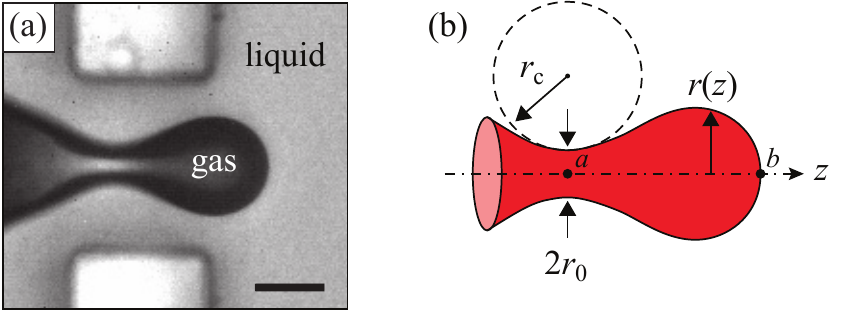}
    \caption{\label{Fig:ICTAM:coordinates}(a) Snapshot of the high-speed recording showing the formation of a microbubble corresponding to frame 86 in Fig.~\ref{Fig:ICTAM:timeseries}. The scale bar denotes 25\,\micron. (b) System of coordinates for an axisymmetric bubble. The shape of the gas-liquid interface $r(z)$ is described as a function of the axial-coordinate $z$. The bubble's volume is the volume enclosed between $a$ and $b$ indicated on the $z$-axis. The gaseous thread forms a neck that is concave in shape with $r_0$ and $r_\mathrm{c}$ the circumferential and axial radius of curvature respectively.}
\end{figure}
The schematic of the axisymmetric shape of the bubble with the axis of symmetry along the $z$-axis is given in Fig.~\ref{Fig:ICTAM:coordinates}.

\begin{figure*}
    \includegraphics[]{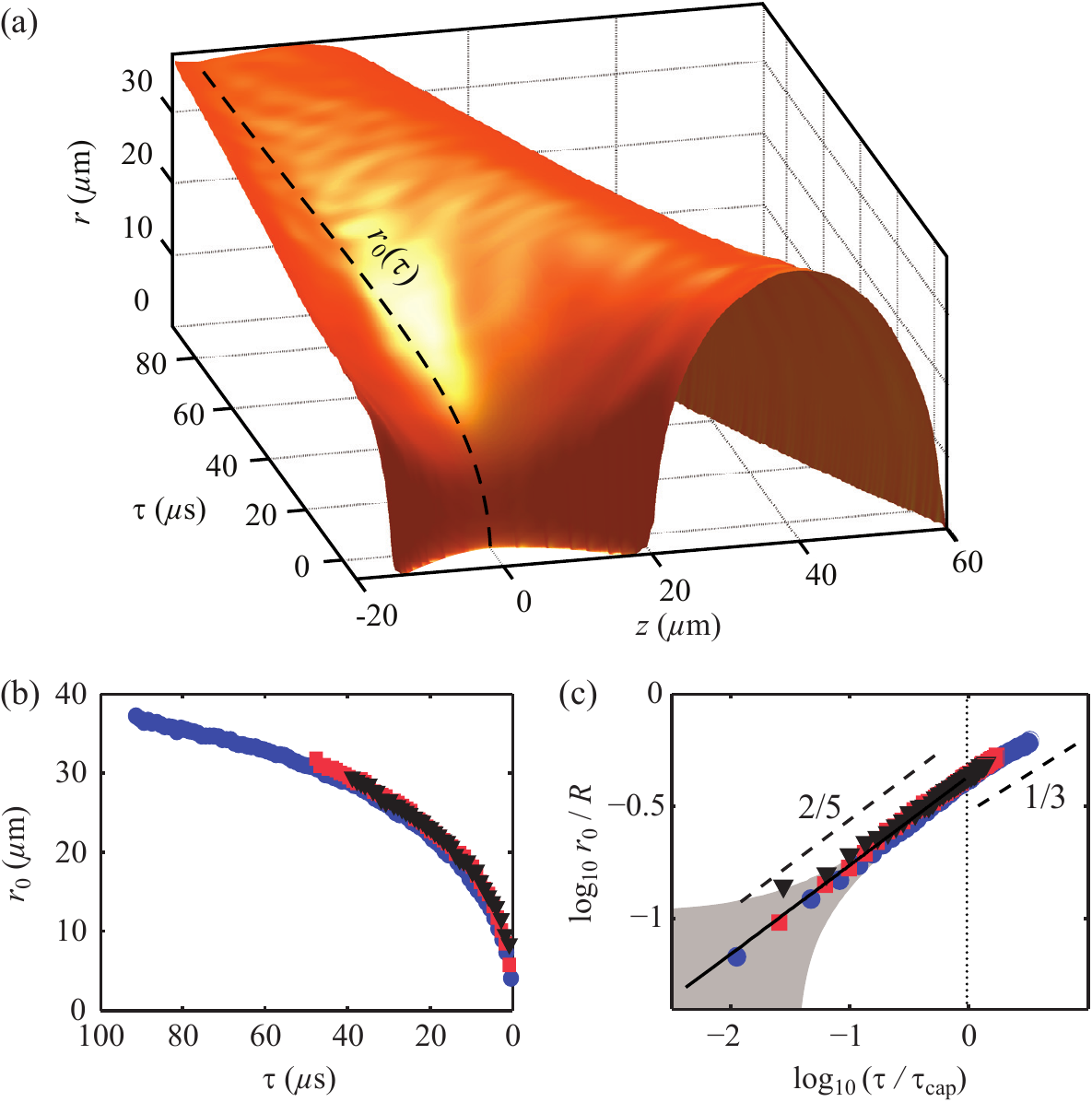}
    \caption{\label{Fig:ICTAM:collapse_curve} (a) Surface contour plot (false color) of the formation of a microbubble. The axisymmetric radius of the bubble is plotted as a function of the axial coordinate $z$ and the time until pinch-off $\tau = t_\mathrm{c} - t$. The dashed line indicates the minimum radius of the neck $r_0$ until final collapse and pinch-off at the origin. (b) The time evolution of the minimum radius of the neck for three different experiments under the same initial conditions. (c) The logarithm of the minimum radius of the neck $r_0$ normalized by the nozzle radius $R = 30$\,\micron as a function of the logarithm of the time until final pinch-off $\tau$, normalized by the capillary time $\tau_\mathrm{cap} = 28$\,\us. The solid line represents the best fit to the data showing a $0.41\,{\pm}\,0.01$ slope. The dashed lines with slope $2/5$ and slope $1/3$ serve as a guide to the eye. The vertical dotted line marks the time closest to pinch-off measured in the work of Dollet \emph{et al.}\cite{Dollet2008} The error in determining the collapse time $t_\mathrm{c}$ is visualized as the gray area.}
\end{figure*}
In Fig.~\ref{Fig:ICTAM:collapse_curve}a a surface contour plot of the radius of the bubble $r$ as a function of the axial coordinate $z$ and the time remaining until pinch-off $\tau$ is shown.
The minimum radius of the neck $r_0$ is indicated by the dashed line.
In Fig.~\ref{Fig:ICTAM:collapse_curve}b we plot $r_0$ as a function of the time remaining until pinch-off $\tau = t_\mathrm{c} - t$, with $t$ the time and $t_\mathrm{c}$ the collapse time, on a linear scale, whereas the collapse curve is represented on a logarithmic scale in Fig.~\ref{Fig:ICTAM:collapse_curve}c.
The collapse time is defined as the moment when the neck reaches its critical radius $r_0 = 0$ and breaks.
From the ultra high-speed imaging results it can be found that this moment occurs between the last frame before actual pinch-off (frame 93 in Fig.~\ref{Fig:ICTAM:timeseries}) and the first frame after pinch-off (frame 94).
We estimate the time of collapse with sub-interframe time accuracy by assuming that the collapse exhibits a power law behavior with $r_0 \propto \left(t_\mathrm{c} - t\right)^\alpha$, where the exponent $\alpha$ and the collapse time $t_\mathrm{c}$ are \emph{a priori} unknown, similarly as was done in Bergmann \emph{et al.}\cite{Bergmann2009}
From a best fit to the data we obtain $t_\mathrm{c} = 93.3\pm{}1$\,\us, where the maximum systematic error is equal to the time between two frames. Note that the error in estimating $t_\mathrm{c}$ results in a deflection of the datapoints away from a straight line bounded between the curves $\log{ r_0 (\tau{}\pm{}1\times{}10^{-6}\,\mbox{s}) / R}$ indicated by the gray area
in Fig.~\ref{Fig:ICTAM:collapse_curve}c.
This figure also suggests that two different stages during bubble formation exist: in the first stage of the collapse, all data was found to be well approximated by a power law $r_0 / R \propto \left( \tau / \tau_\mathrm{cap} \right)^\alpha$, with $\alpha = 0.29\,{\pm}\,0.02$. In the final stage, when $\tau \le \tau_\mathrm{cap}$, a scaling with exponent $\alpha = 0.41\,{\pm}\,0.01$ is observed, spanning almost two decades.

\subsection{Liquid inertia driven pinch-off}
Approaching the singularity at pinch-off ($\tau \rightarrow 0$), the relative importance between viscous forces, surface tension forces, and inertial forces are given by the Reynolds number, Weber number, and the capillary number.
The Reynolds number, as a measure of the ratio between inertial forces to viscous forces, is expressed as
\begin{equation}
    \textrm{Re} = \frac{\rho r_0 \dot{r}_0}{\eta}, 
\end{equation}
with characteristic length scale $r_0$, and characteristic velocity equivalent to the radial velocity of the interface $\dot{r}_0$. 
The relative importance of inertial forces with respect to surface tension forces is given by the Weber number
\begin{equation}
    \textrm{We} = \frac{\rho r_0 \dot{r}_0^2}{\gamma}\,. 
\end{equation}
The capillary number represents the relative importance of viscous forces to surface tension forces as
\begin{equation}
    \textrm{Ca} = \frac{\eta \dot{r}_0}{\gamma}\,.
\end{equation}
If we now assume that $r_0 \propto \tau^\alpha$, where the experimentally determined scaling exponent is close to $\alpha = 2/5$, it follows that $\textrm{Re} \propto \tau^{-1/5}$, $\textrm{We} \propto \tau^{-4/5}$, and $\textrm{Ca} \propto \tau^{-3/5}$.
This implies that $\textrm{Re}$, $\textrm{We}$, and $\textrm{Ca}$ all diverge approaching the singularity. Accordingly, inertial forces must dominate both surface tension and viscous forces, hence the final stage of the collapse is purely liquid inertia dominated.

\begin{figure}
    \includegraphics[]{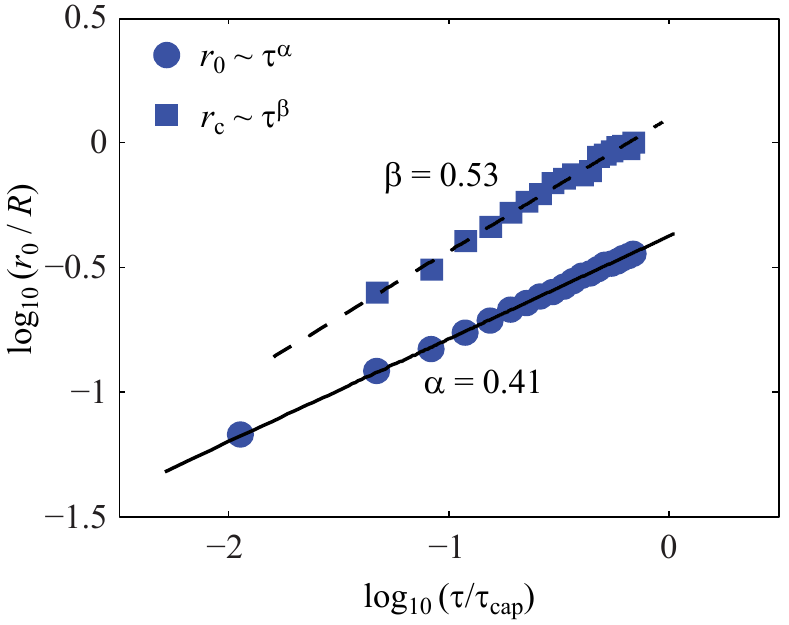}
    \caption{\label{Fig:ICTAM:slenderness}The axial radius of curvature $r_\mathrm{c}$ (squares) decreases faster compared to the circumferential radius of curvature $r_0$ (bullets). Hence the neck becomes less slender approaching pinch-off. Thus the slenderness ratio $\lambda = r_\mathrm{c} / r_0$, becomes smaller when approaching the pinch-off. The radii of curvature are normalized by the nozzle radius $R = 30$\,\micron{}. The time until pinch-off is normalized by the capillary time $\tau_\mathrm{cap} \approx 28$\,\us.}
\end{figure}
In Eggers \emph{et al.}\cite{Eggers2007} it was shown that for a liquid inertia driven collapse both the radial and axial length scale of the neck are important. Hence, the time evolution of the shape of the neck is investigated by measuring its slenderness.
The slenderness ratio $\lambda$ is defined as the ratio of the axial radius of curvature to the circumferential radius of curvature of the neck. The larger the slenderness ratio is, the more slender the neck is.
The axial radius of curvature is measured by locally fitting a circle with radius $r_\mathrm{c}$ to the contour of the neck, whereas the circumferential radius of curvature $r_0$ is equivalent to the minimum radius of the neck, see Fig.~\ref{Fig:ICTAM:coordinates}.
In Fig.~\ref{Fig:ICTAM:slenderness} the time evolution of the principal radii of curvature are plotted on a logarithmic scale for the final stage of the collapse.
It is found that the axial radius of curvature exhibits a power law behavior $r_\mathrm{c} \propto \tau^{\beta}$, with $\beta = 0.53$. The circumferential radius of curvature scales as $r_0 \propto \tau^\alpha$, with $\alpha = 0.41$, as was shown before (\emph{cf.}~Fig.~\ref{Fig:ICTAM:collapse_curve}c).

The axial radius of curvature is found to have the more rapidly diminishing exponent ($\beta > \alpha$), which implies that the slenderness $\lambda = r_\mathrm{c} / r_0 \propto \tau^{\beta} / \tau^\alpha \rightarrow 0$ for $\tau \rightarrow 0$. In other words, the neck profile becomes \emph{less slender} approaching pinch-off, thus, both the radial and the axial length scales are still important.
This 3D character implies that the liquid flows spherically inward towards the collapsing neck.
Thus, it might be anticipated that, this 3D collapse can be approximately described using the Rayleigh-Plesset equation for spherical bubble collapse\cite{Oguz1993}
\begin{equation}
    r_0 \ddot{r}_0 + \frac{3}{2} \dot{r}_0^2 = \frac{1}{\rho} \left(p - \frac{2 \gamma}{r_0}\right),
\end{equation}
with capillary pressure $p$. It should be noted that a necessary condition for this is that the neck should be much smaller than the channel dimensions ($r_0 \ll W$, $H$).
By substituting $r_0 \propto \tau^\alpha$ in above equation and keeping the right-hand side constant, which assumes an inertia-dominated flow, it is found that it is necessary that $\alpha = 2/5$, which agrees surprisingly well with our experimental findings of $0.41\,\pm\,0.01$.

\subsection{``Filling effect''}
How to account for the scaling $r_0 \propto \tau^{0.29 \pm 0.02}$ for $\tau > \tau_\mathrm{cap}$, \emph{i.e.}~at early times? At this initial stage of the collapse a thin layer of liquid with a thickness of several micrometers separates the bubble from the hydrophilic channel wall.\cite{Garstecki2004} The liquid flow in such a confined channel can be described using Darcy's law for pressure driven flow through porous media. The volumetric flow rate of liquid that permeates into the neck region is
\begin{equation}
    \label{eq:liquidflow}
    Q_\mathrm{in} = -\frac{k A}{\eta} \frac{\partial{}p}{\partial{}z},
\end{equation}
with $k$ the permeability, $A = W H - \pi r_0^2$ the cross-sectional area of the thin liquid layer surrounding the bubble, and $\partial{}p / \partial{}z$ the pressure gradient.

The pressure distribution in the liquid is inhomogeneous, thus the bubble's surface does not have a constant curvature even though the gas pressure is practically uniform. The pressure gradient that drives the liquid flow can be derived from the capillary pressure
\begin{equation}
    \label{eq:capillarypressure}
    p = \gamma \left( \frac{1}{r_0} - \frac{1}{r_\mathrm{c}} \right).
\end{equation}
In the initial stage of the collapse, \emph{i.e.}~at the onset of neck formation, $r_\mathrm{c} > r_0$. As a gross simplification, we approximate the neck as a radially collapsing cylinder, of length $r_\mathrm{c}$ much larger than its radius $r_0$. The capillary pressure is then $p \approx \gamma / r_0$, therefore $\partial{}p/\partial{}z \approx -r_0^{-2} \partial{}r_0/\partial{}z$.

The volumetric gas flow rate that is pushed out of the neck region is
\begin{equation}
    \label{eq:gasflow}
    Q_\mathrm{out} = -\dot{V}_\mathrm{g} \approx -r_0 r_\mathrm{c} \dot{r}_0,
\end{equation}
with $V_\mathrm{g} \approx r_\mathrm{c} r_0^2$ the volume occupied by the gas.

The gas in the neck is replaced by the liquid. This is referred to as the ``filling effect'', thus, from a balance between Eq.~(\ref{eq:liquidflow}) and Eq.~(\ref{eq:gasflow}), we now get
\begin{equation}
    r_0 r_\mathrm{c} \dot{r}_0 \approx \frac{1}{r_0^2} \frac{\partial{}r_0}{\partial{}z} \approx \frac{1}{r_0 r_\mathrm{c}},
\end{equation}
hence, assuming that $r_\mathrm{c}$ varies little in this initial stage of the collapse, $r_0^2 \dot{r}_0$ is roughly constant. It follows that the radius of the neck must scale as $r_0 \propto \tau^\alpha$, with $\alpha = 1/3$. This is in good agreement with the experimentally measured scaling exponent $\alpha \approx 0.29\,{\pm}\,0.02$ for $\tau > \tau_\mathrm{cap}$.

\section{Discussion}
In Gekle \emph{et al.}\cite{Gekle2010} a supersonic air flow through the neck is visualized using smoke particles and it is reported that Bernoulli suction accelerates the collapse. An accelerated collapse due to Bernoulli suction is also reported by Gordillo \emph{et al.},\cite{Gordillo2005} giving rise to a 1/3 scaling exponent. It is extremely difficult to measure the gas velocity in a microfluidic flow-focusing device in a direct way. However, the camera's wide field of view (200\,\micron{}$\,{\times}\,$175\,\micron) enabled us to capture the contour of the expanding bubble in great detail and allows for an estimate of the gas velocity.

\begin{figure}
    \includegraphics[width=8cm]{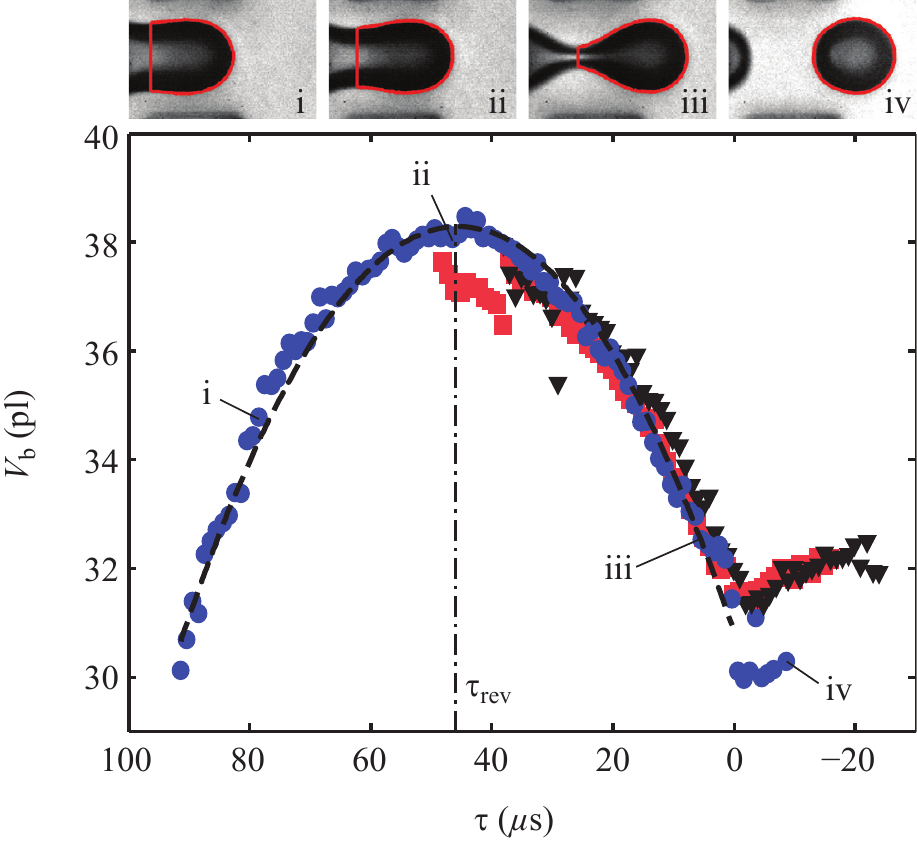}
    \caption{\label{Fig:ICTAM:volume}Volume of the bubble $V_\mathrm{b}$ as a function of the time until pinch-off $\tau$. The volume of the bubble was calculated by integration over its contour along the $z$-axis between the neck and the tip of the bubble (indicated $a$ and $b$ in Fig.~\ref{Fig:ICTAM:coordinates}). The insets (i--iv) depict the contours that enclose the bubble's volume corresponding to the marked data points in the graph. A second order polynomial fit is used to calculate $\partial{}{V}_\mathrm{b}/\partial{}\tau$ (dashed line). The bubble reaches its maximum volume at $\tau_\mathrm{rev}$ (dashed-dotted line) and the gas flow direction reverses, consequently, the bubble shrinks during the final moments before pinch-off. Different symbols represent different experiments, giving an indication of the reproducibility.}
\end{figure}
The volume of the bubble $V_\mathrm{b}$, as the gas volume downstream of the neck that is enclosed by the bubble contours, is calculated as follows:
\begin{equation}
    V_\mathrm{b} = \int_a^b \textrm{d}z \pi{} r^2(z),
\end{equation}
with the profile of the bubble $r(z)$, with $a$ the axial coordinate of the location of the neck and $b$ the tip of the bubble (\emph{cf.}~Fig.~\ref{Fig:ICTAM:coordinates}).
We plot the bubble volume $V_\mathrm{b}$ as a function of time until pinch-off in Fig.~\ref{Fig:ICTAM:volume}.
The bubble's contour is indicated for four characteristic moments during the bubble formation process in the panels (i--iv) of Fig.~\ref{Fig:ICTAM:volume}.
In the initial stage the gaseous thread in front of the flow-focusing channel is forced to enter the channel and completely fills it (i--ii). The volume of the bubble increases until it reaches a maximum volume of 38\,pl at $\tau = 46$\,\us (ii). The restricted liquid flow starts to ``squeeze'' the gaseous thread and a clearly visible neck begins to develop. Then a remarkable event takes place---the gas flow reverses and the neck starts to collapse (ii-iii) until bubble pinch-off occurs. The volume of the bubble beyond pinch-off ($\tau < 0$) is 31\,pl, which is equivalent to a bubble radius of 19\,\micron{}.

We estimate the volumetric gas flow rate $Q_\mathrm{g}$ through the neck as the time derivative of the volume of the bubble, $Q_\mathrm{g} = \dot{V}_\mathrm{b}$, where it is assumed that no gas diffusion into the surrounding liquid take place. The bubble volume is approximated by a second order polynomial function, as indicated by the dashed line in Fig.~\ref{Fig:ICTAM:volume}, which is used to obtain the time derivative of the volume.

\begin{figure}

    \includegraphics[]{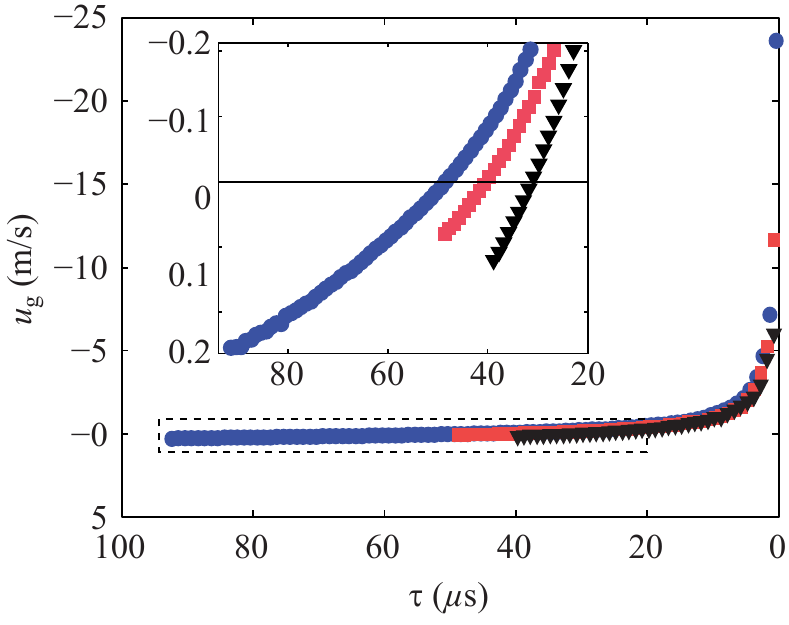}
    \caption{\label{Fig:ICTAM:velocity}Gas velocity $u_\mathrm{g}$ in the neck as a function of the time until collapse $\tau$. The initial positive gas velocity reverses its flow direction and accelerates when approaching the pinch-off. At the moment of pinch-off maximum velocity of $-23$~m/s is reached. In the inset an enlarged section of the graph for the data points encircled by the dashed line is represented demonstrating the gas flow reversal. Again, different symbols/colors represent different individual experiments.}
\end{figure}
The gas velocity through the neck $u_\mathrm{g}$ is calculated as the volume flow rate $Q_\mathrm{g}$ divided by the cross-sectional area of the neck ($\pi r_0^2$). In Fig.~\ref{Fig:ICTAM:velocity} we plot the gas velocity as a function of the time until pinch-off.
Note that the gas velocity is low during almost the entire collapse process, \emph{i.e.}~$\left| u_\mathrm{g} \right| < 0.5$\,m/s (see the inset in Fig.~\ref{Fig:ICTAM:velocity}), however, in the final stage of the collapse, a strong acceleration of the gas flow is observed, reaching a velocity up to $u_\mathrm{max} = -23$\,m/s.
The pressure drop due to the fast gas flow through the neck is given by Bernoulli's equation as $\Delta{}p = -\rho_\mathrm{g} \left(u_\mathrm{rev}^2 - u_\mathrm{max}^2\right) / 2 \approx 0.3$\,kPa, with  $u_\mathrm{rev} = 0$\,m/s the gas velocity at the moment of flow reversal and $\rho_\mathrm{g} = 1.2$\,kg/m$^3$ the gas density.

\begin{figure}
    \includegraphics[]{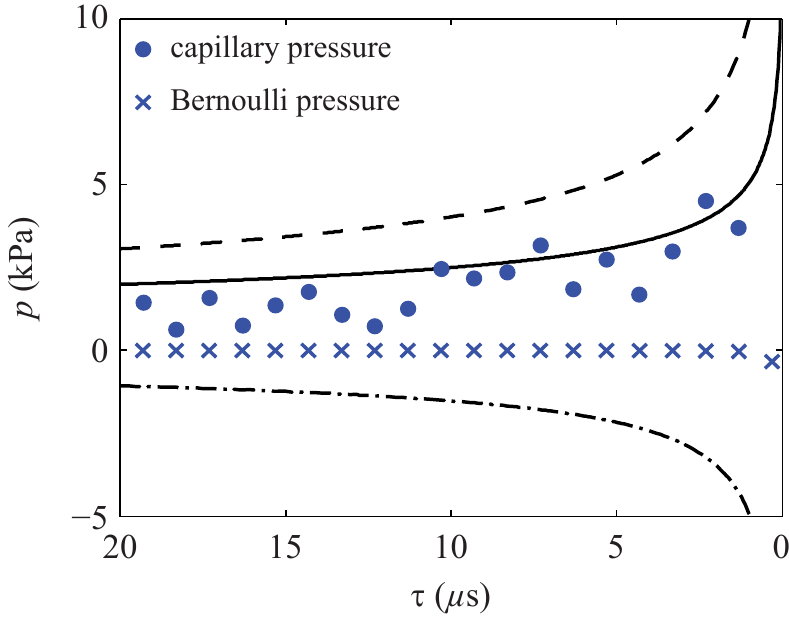}
    \caption{\label{Fig:ICTAM:pressures}Evolution of the capillary pressure and the Bernoulli pressure drop during the final moments of bubble pinch-off. The dashed line indicates the capillary pressure contribution due to the circumferential curvature ($p = \gamma / r_0$, with $r_0 \propto \tau^{0.41}$); the dashed-dotted line shows the pressure contribution from the axial curvature ($p = - \gamma / r_\mathrm{c}$, with $r_\mathrm{c} \propto \tau^{0.53}$); the solid line represents the sum of both contributions. The bullets represents the capillary pressure obtained from the local shape of the neck ($r(z)$) extracted from the wide-field-of-view images and using Eq.~(\ref{capillary_pressure_axisymmetric}). The increasing gas velocity through the neck causes a local pressure reduction in the neck, referred to as Bernoulli suction ($p = -\rho u^2_\mathrm{g}/2$), as indicated by the crosses. It can be seen that the capillary pressure clearly dominates during the entire bubble formation process.}
\end{figure}
We now compare this pressure drop with the capillary pressure in the neck.
Just before pinch-off, the capillary pressure, as a consequence of surface tension forces acting on the curved interface, should be at a much higher pressure than the surrounding liquid.
For an axisymmetric surface profile, with $r = r(z)$, the capillary pressure, as a function of the axial coordinate $z$, is given by the Laplace equation 
\begin{equation}
    \label{capillary_pressure_axisymmetric}
    p(z) = \gamma \left\lbrack \frac{1}{r \sqrt{1 + {r^\prime}^2}} - \frac{r^{\prime\prime}}{\left(1 + {r^\prime}^2\right)^{3 / 2}} \right \rbrack,
\end{equation}
where prime denotes the derivative with respect to $z$.\cite{deGennes}
Note that, at the location where the neck is thinnest, the first term equals the circumferential curvature ($r_0^{-1}$), whereas the second term represents the axial curvature ($r_\mathrm{c}^{-1}$).
In Fig.~\ref{Fig:ICTAM:pressures}, both the Bernoulli pressure drop and the capillary pressure in the neck are plotted as a function of the time remaining until pinch-off.
The capillary pressure (represented by the bullets) is obtained by inserting the complex shape of the neck into Eq.~(\ref{capillary_pressure_axisymmetric}).
The dashed line and the dashed-dotted line indicate the pressure contribution from the circumferential curvature ($p \propto \gamma \tau^{-\alpha}$) and the axial curvature ($p \propto -\gamma \tau^{-\beta}$) respectively, while the solid line represents the sum of both contributions of the capillary pressure.
In the figure it is demonstrated that the pressure drop due to Bernoulli suction is marginal in comparison to the increasing capillary pressure approaching pinch-off. Hence, it can be concluded that Bernoulli suction, \emph{i.e.}~gas inertia, is irrelevant during the entire bubble formation process.
It is also shown that the concave axial curvature counteracts the circumferential curvature leading to a significant decrease in capillary pressure. This confirms that the axial length scale of the neck is important and gives the collapse a three-dimensional character.

\section{Conclusion}
In conclusion, we visualized the complete microbubble formation and extremely fast bubble pinch-off in a microscopically narrow flow-focusing channel of square cross-section ($W\,{\times}\,H = 60$\,\micron{}\,${\times}\,60$\,\micron{}), using ultra high-speed imaging. The camera's wide field of view enabled visualization of all the features of bubble formation, including the two principal radii of curvature of the bubble's neck. Recording was performed at 1\,Mfps, thereby, approaching the moment of pinch-off to within 1\,\us. It was found that the neck's axial length scale decreases faster than the radial one, ensuring that the neck becomes less and less slender, collapsing spherically towards a point sink. 
We describe this collapse using the Rayleigh-Plesset equation for spherical bubble collapse,\cite{Oguz1993} and recover a $2/5$ power law exponent which is consistent with our experimental findings.
The gas velocity through the neck is calculated from the growth-rate of the bubble. Just before pinch-off the gas velocity accelerates up to $-23$~m/s reducing the bubble's volume, however this velocity is too low for Bernoulli suction to be the dominant effect.
Thus, the final moment of microbubble pinch-off in a flow-focusing system is purely liquid inertia driven.

\section{Acknowledgement}
We kindly acknowledge J.\,M.\ Gordillo for insightful discussions.
This work was financially supported by the MicroNed technology program of the Dutch Ministry of Economic Affairs through its agency SenterNovem under grant \mbox{Bsik-03029.5}.


\end{document}